%% file: main.tex
\documentclass[conference]{IEEEtran}
\IEEEoverridecommandlockouts

\usepackage{cite}
\usepackage{amsmath,amssymb,amsfonts}
\usepackage{graphicx}
\usepackage{textcomp}
\usepackage{xcolor}
\usepackage[hyphens]{url}

\usepackage{enumitem}
\usepackage{pifont}

\usepackage[ruled,vlined,linesnumbered]{algorithm2e}
\usepackage{tikz}

\newcommand{\colimgwidth}{\columnwidth}
\newcommand{\fullimgwidth}{\textwidth}

\newcommand{\func}[1]{\textnormal{\textsc{#1}}}

\newcommand*\myfilledcircled[1]{\,\tikz[baseline=(char.base)]{
    \node[shape=circle, fill=black, text=white, inner sep=0.4pt, font=\small] (char) {#1};}}

\usepackage{xcolor}
\definecolor{myNewColor}{HTML}{0055AA}

\def\BibTeX{{\rm B\kern-.05em{\sc i\kern-.025em b}\kern-.08em
    T\kern-.1667em\lower.7ex\hbox{E}\kern-.125emX}}
\begin{document}

\pdfpagewidth=8.5in
\pdfpageheight=11in

\newcommand{\iscasubmissionnumber}{1216}


\pagenumbering{arabic}

\title{Cassandra: Enabling Reasoning LLMs at Edge \\ via Self-Speculative Decoding}

\author{\IEEEauthorblockN{Soongyu Choi, Yuntae Kim, Muyoung Son, and Joo-Young Kim}
\textit{KAIST}\\
Daejeon, Republic of Korea \\
\{soongyu1291, yuntaekim, kkt1690, jooyoung1203\}@kaist.ac.kr}


\maketitle
\thispagestyle{plain}
\pagestyle{plain}


\input{Contents/0_Abstract}


\input{Contents/1_Introduction}

\input{Contents/2_Background}

\input{Contents/3_Motivation}

\input{Contents/4_Speculative_Decoding_Algorithm}
\input{Contents/5_Hardware_Architecture}

\input{Contents/6_Methodology_and_Evaluation}

\input{Contents/7_Analysis_and_Discussion}

\input{Contents/8_Related_Work}

\input{Contents/9_Conclusion}

\section*{Acknowledgment}
This work was supported by Institute for Information \& communications Technology Promotion (IITP) grant funded by the Korea government (MSIT) (No. RS-2025-02264029, Integration and Validation of an AI Semiconductor-Based Data Center Training and Inference System and No. RS-2023-00228255, PIM-NPU Based Processing System Software Developments for Hyper-scale Artificial Neural Network Processing).


\bibliographystyle{IEEEtranS}
\bibliography{refs}

\end{document}

%% file: Contents/0_Abstract.tex
\begin{abstract}

Speculative decoding has emerged as a promising lossless approach to accelerating Large Language Models (LLMs). Recently, as the overhead of the decode stage has increased in reasoning LLMs, while algorithmic approximations cause significant accuracy degradation, lossless acceleration through speculative decoding has become a key component for efficient LLM serving systems. However, despite numerous advances, a speculative decoding method that can provide sufficient performance at low batch sizes without requiring additional training remains elusive. This presents a considerable challenge for achieving lossless acceleration in low-batch LLM inference on consumer-grade devices.

To address this limitation, we propose Cassandra, an algorithm–hardware co-designed self-speculative decoding framework tailored explicitly for low-batch scenarios. The core idea of Cassandra is to construct a high-performance, training-free draft model through fine-grained data selection. By leveraging optimized pruning and mantissa truncation, Cassandra identifies and isolates the most salient values within both model weights and the Key–Value (KV) cache. These selected values are first loaded to rapidly generate candidate tokens, followed by parallel verification using full-precision data. This design achieves substantial performance gains over prior self-speculative decoding methods, which primarily rely on layer skipping or structured KV cache compression. In addition, to mitigate the overhead of format conversion between the Cassandra representation and standard floating-point formats, we introduce a lightweight encoder–decoder hardware module designed for seamless integration with commercial GPUs and NPUs.

Our evaluation demonstrates that Cassandra achieves a performance improvement of up to 2.41$\times$ than the BFloat16 baseline without additional training. Moreover, when running Llama3-8B on an RTX 4090, Cassandra is capable of generating 1.81$\times$ more tokens under a fixed memory budget compared to Eagle-3, a state-of-the-art speculative decoding method.

\end{abstract}

%% file: Contents/1_Introduction.tex
\section{Introduction}
\label{section_1}

The emergence of ChatGPT ushered us into the irreversible era of AI. Today, major Large Language Model (LLM) services serve hundreds of millions of monthly active users, driving an unprecedented demand for efficient inference systems. As a result, designing high-performance LLM inference infrastructure has become one of the central challenges in the computer architecture community.

LLM inference can be broadly divided into two distinct phases: prefill and decode. The prefill stage processes the input prompt and requires the simultaneous computation of a large number of tokens, thereby demanding high computational performance. In contrast, the decode stage operates in an autoregressive manner, generating one token at a time. By reusing intermediate Key–Value (KV) tensors from previous steps, this phase typically exhibits low arithmetic intensity and limited parallelism.

On modern xPU platforms (e.g., GPUs and NPUs), the decode stage often becomes the primary performance bottleneck due to its memory-bound nature. To address this challenge, prior work has largely focused on two complementary strategies. The first is input batching, which aggregates multiple user requests to increase effective arithmetic intensity during decoding~\cite{oaken, attacc, orca, duplex, splitwise}. By improving hardware utilization, batching enables higher throughput through concurrent token generation. The second approach involves algorithmic optimizations such as quantization and data pruning—collectively referred to as lossy compression~\cite{amove, mx+, tender, microscopiq, qserve, duquant}. While such techniques inherently introduce some degree of accuracy degradation, carefully designed methods can substantially improve performance by reducing both compute requirements and memory bandwidth, while keeping accuracy loss minimal.




However, two emerging trends in the LLM ecosystem introduce new challenges. The first is the increasing demand for small LLM inference on consumer-grade devices. Traditionally, both academia and industry have focused on datacenter-scale inference systems. Recently, however, open source small LLM models~\cite{Llama3-8B, Qwen3-4B-Thinking-2507, MobileLLM-R1-950M, Gemma3-270m} have demonstrated competitive performance, bringing LLM inference on consumer-grade devices into the spotlight.

The second trend is the rise of reasoning LLMs~\cite{Qwen3-8B, Deepseek-R1-Distillated-Llama3-8B}. These models, typically enhanced via reinforcement learning on high-quality datasets, exhibit strong capabilities in complex tasks such as mathematical reasoning and code generation. Notably, reasoning capabilities are no longer limited to large models; even relatively small models have recently begun to incorporate such functionality~\cite{Qwen3-4B-Thinking-2507, MobileLLM-R1-950M}.

These trends jointly expose a critical challenge in edge deployment. Reasoning workloads often produce significantly longer output sequences, causing the decode stage to dominate end-to-end latency. At the same time, batching—one of the most effective optimizations in datacenter environments—is largely infeasible on edge devices, where workloads typically involve a single or a small number of concurrent users. As a result, optimization opportunities are effectively limited to algorithmic approaches such as lossy compression.

The problem is that lossy compression suffers from severe performance degradation when applied to reasoning LLMs. Considering that even very small LLMs now support reasoning capabilities, this issue acts as a critical bottleneck that undermines the potential of LLMs on consumer-grade hardware.



To address this challenge, we argue that speculative decoding offers a promising direction for high-accuracy acceleration in edge environments. Speculative decoding improves decode-stage efficiency by leveraging a lightweight draft model to predict multiple candidate tokens, which are then verified in parallel. This approach enables significant speedup while preserving output quality.

Despite its potential, existing speculative decoding methods face several limitations, including high training cost, limited effectiveness in low-batch settings, additional memory overhead, and diminishing returns for long output sequences. These drawbacks hinder their applicability to edge scenarios. To overcome these challenges, we propose Cassandra, a hardware–software co-designed speculative decoding framework tailored for low-batch inference on consumer-grade devices. Our main contributions are as follows:

\begin{enumerate}[label={(\arabic*)}] 
    \item 
        We introduce Cassandra, an algorithm–hardware co-designed speculative decoding method. Our key insight is that a compact yet high-quality draft model can be derived directly from the target LLM through fine-grained data selection, without additional training. Based on this observation, Cassandra applies unstructured value pruning and mantissa truncation to the original model, extracting the most salient information at the bit level to construct the draft model. Furthermore, we identified that the large size of the exponent becomes a bottleneck, preventing the performance improvement in this scenario. As two distinct solutions, we propose lossy exponent compression based on the MX format and lossless exponent compression based on unary coding. By incorporating the fine-grained draft model and exponent compression, Cassandra achieves a performance improvement of up to 2.41$\times$ compared to BFloat16 baseline, which is also significantly higher compared to existing methods.
    \item
        In Cassandra, weights and KV cache are stored in a specialized format obtained through unstructured pruning, mantissa truncation, and exponent compression. To execute computation on commercial xPU architectures, these representations must be converted back to standard floating-point format. To prevent this conversion from becoming a performance bottleneck, we propose a lightweight Cassandra encoder–decoder. The proposed design supports both MX-based and unary-based exponent compression within a unified architecture, enabling efficient format conversion with minimal power and area overhead. Furthermore, we provide integration guidelines for incorporating the encoder–decoder into conventional GPU and NPU architectures, along with a memory management scheme that mitigates irregular memory access patterns.
    
    \item 
        To the best of our knowledge, Cassandra is the first self-speculative decoding framework that leverages algorithm–hardware co-design specifically for low-batch LLM inference on consumer-grade edge devices.
        
\end{enumerate}

%% file: Contents/2_Background.tex
\section{Background}
\label{section_2}

\begin{figure}[t]
\centering
\includegraphics[width=\colimgwidth]{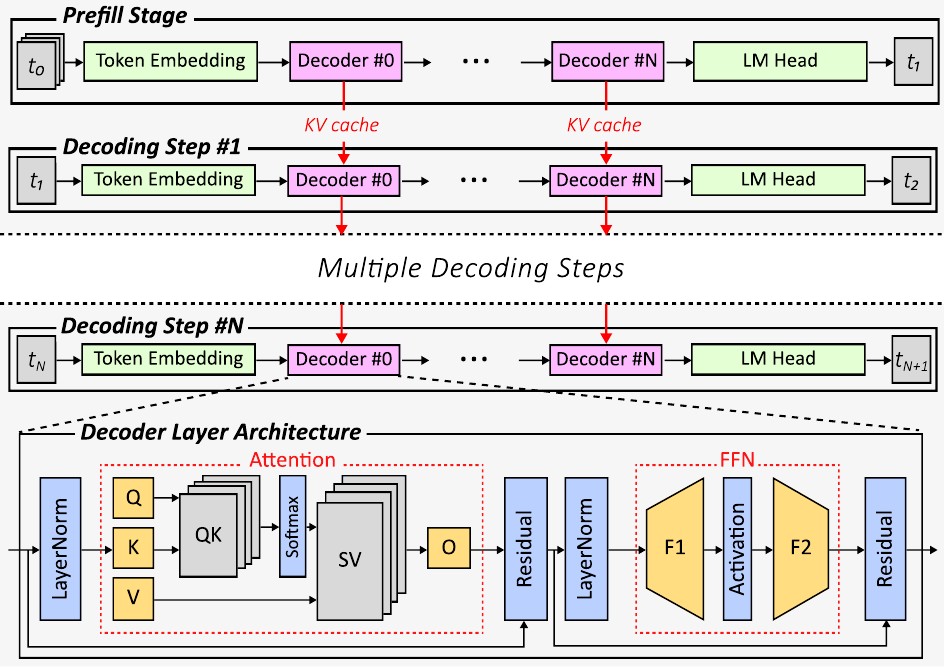}
\caption{Architecture of autoregressive transformer based LLMs}
\label{fig_2_1}
\vspace{-1.5em}
\end{figure}

\subsection{Large Language Model Architecture}
\label{section_2_1}
Despite the exploration of alternative language model architectures~\cite{rnn-rwkv, mamba, llada}, the autoregressive transformer remains the dominant architecture for LLMs. Transformer fundamentally consists of two main layers: the multi-head attention and the feed-forward network. Figure~\ref{fig_2_1} illustrates the structure of a transformer layer, where the yellow modules denote weight–activation multiplications, the gray modules represent activation–activation multiplications, and the blue modules correspond to non-linear operations.

A defining characteristic of transformers in LLMs is their autoregressive nature. These models are trained to predict the next token conditioned on the preceding sequence. Consequently, text generation is performed through an autoregressive process, in which the entire accumulated sequence is repeatedly processed to produce each subsequent token. A naive implementation of this process incurs substantial redundant computation, as portions of the sequence remain unchanged across generation steps.

To mitigate this inefficiency, modern LLMs employ the Key-Value (KV) caching technique, which stores the key and value tensors generated by the attention layers and reuses them in subsequent iterations. The initial phase, where the model processes the input sequence and generates the first token, is referred to as the prefill stage. The subsequent token generation steps that leverage the KV cache constitute the decode stage. The prefill stage is dominated by large-scale matrix multiplications, resulting in high arithmetic intensity and substantial computational demand. In contrast, the decode stage exhibits significantly lower arithmetic intensity and is typically memory-bound on conventional parallel architectures.

\subsection{Speculative Decoding}
\label{section_2_3}

\begin{figure}[t]
\centering
\includegraphics[width=\colimgwidth]{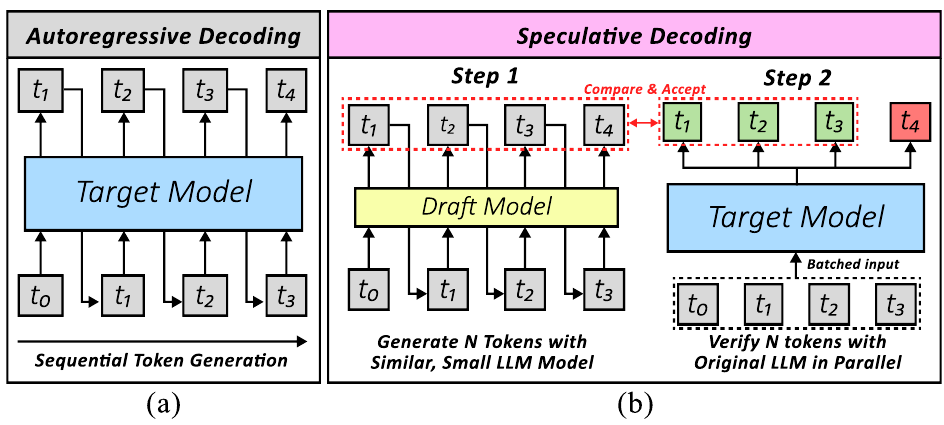}
\caption{(a) Autoregressive decoding. (b) speculative decoding.}
\label{fig_2_3}
\vspace{-1.5em}
\end{figure}

The core idea of speculative decoding is to transform the sequential token generation in the decode stage into parallel ones. Based on this idea, a wide variety of parallel decoding techniques have recently emerged. Still, the method that uses two LLMs, a target model and a draft model, remains the most widely adopted~\cite{speculativedecoding}.

Figure~\ref{fig_2_3} illustrates the difference between autoregressive decoding and speculative decoding. Here, the target model is the original LLM, and the draft model is a comparatively smaller LLM trained to generate tokens with a probability distribution similar to the target model. In this method, the draft model first generates a few tokens, and the target model subsequently verifies those tokens in parallel. To understand the performance gain afforded by speculative decoding through a simplified example, let us define the time required for the draft model to generate a single token as $t_{draft}$, and the time required for the target model to generate a single token as $t_{target}$. In a scenario involving $\gamma$ token generations via autoregressive decoding, the total time consumption would be $\gamma t_{target}$. However, consider the application of speculative decoding, where $\gamma$ tokens are first generated by the draft model and subsequently verified by the target model, assuming all $\gamma$ tokens are generated correctly. In this case, the target model performs batched inference by receiving $\gamma$ tokens from the draft model. Since $\gamma$ is generally not large, this batched inference operates as a memory-bound on commercial xPUs, and the time required for this is nearly equivalent to the time needed to generate a single token. Consequently, the total time taken in this scenario is approximated by $\gamma t_{draft}+t_{target}$. If $t_{draft}$ is sufficiently smaller than $t_{target}$, a substantial performance improvement is achieved.


One problem arises in this process: the draft model does not always generate the correct tokens. If the draft model generates an incorrect token, the target model detects it during the verification process and only accepts tokens up to the one preceding the error as correctly generated. The criteria for accepting tokens generated by the draft model can vary depending on the scenario.

When utilizing greedy decoding, we can only accept the output if the draft model produces the identical token. Under this condition, the target model's output remains perfectly preserved by speculative decoding. In contrast, recent LLMs often employ sampling techniques that introduce a degree of randomness during the new token generation process.
In such a scenario, to maintain appropriate randomness, a probability-based sampling method~\cite{speculativesampling}, often expressed by the equation below, is frequently employed.

\begin{equation}
n \leftarrow \min\left(\left\{i-1 \middle| 1 \leq i \leq \gamma, r_i > \frac{p_i(x)}{q_i(x)}\right\} \cup \{\gamma\}\right)
\end{equation}
\vspace{0.05in}

Here, $n$ is the number of accepted tokens, $\gamma$ is the draft length, $i$ is the index, $r_i$ is a random probability between 0 and 1, $p_i(x)$ is the probability of the $i$-th token obtained from the target model's logits, and $q_i(x)$ is the probability of the $i$-th token obtained from the draft model's logits. This approach is known as rejection sampling, and with this method, the entire speculative decoding system is mathematically guaranteed to generate tokens that are probabilistically identical to those of the target model.

The ratio in which the draft model correctly predicts how many tokens on average is called the acceptance rate. Since a low acceptance rate for the draft model can lead to a performance degradation in speculative decoding, it is crucial to develop a draft model that is both fast and has a sufficiently high acceptance rate.


%% file: Contents/3_Motivation.tex
\section{Motivation}
\label{section_3}

\begin{figure}[t]
\centering
\includegraphics[width=\colimgwidth]{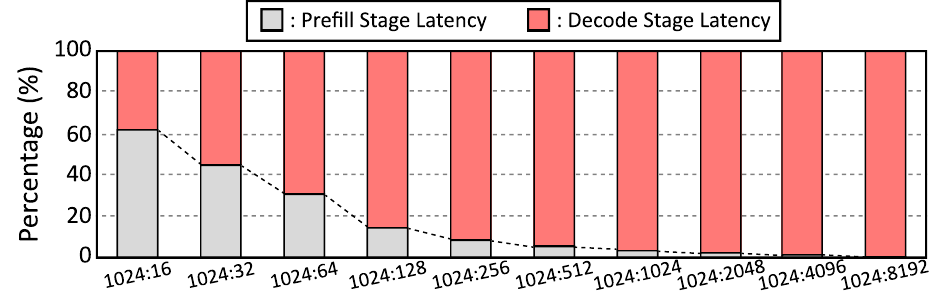}
\caption{A latency ratio of prefill stage and decode stage in single batch Llama3-8B inference on RTX 4090. Total is 100\% and the X-axis means the number of input tokens: output tokens}
\label{fig_3_1}
\vspace{-0.5em}
\end{figure}
\subsection{Decode stage Overhead in xPU}
\label{section_3_1}

Both the prefill and decode stages must be carefully considered in the design of LLM inference systems. However, once a system is optimized for the prefill stage—typically through a highly parallel architecture—the performance bottleneck shifts to the decode stage. Most commercial GPUs and NPUs fall into this category. Figure~\ref{fig_3_1} illustrates the ratio of prefill and decode stage latencies according to token lengths on an Nvidia RTX 4090~\cite{RTX4090}. As shown in the figure, even when the number of prefill tokens and decoding tokens is identical, the decode stage accounts for 98\% of the end-to-end latency.

This imbalance is further exacerbated by the recent emergence of reasoning LLMs. Such models tend to generate significantly longer output sequences compared to conventional LLMs. Table~\ref{table_2_1} presents the context lengths produced on the GPQA-Diamond~\cite{GPQA-Diamond} benchmark by two models sharing the same base architecture. As shown, reasoning LLMs generate sequences that are approximately 2.99$\times$ to 5.20$\times$ longer than their non-reasoning counterparts. As a result, the decode stage becomes substantially prolonged, further reinforcing its dominance in overall latency.

This outcome may change in cases of large-batch inference. Nevertheless, in the case of low-batch LLM inference acceleration, prioritizing performance improvement in the decode stage is more appropriate than enhancing computational capabilities.


\begin{table}[t]
    \caption{Context length comparison of various LLMs}
    \vspace{-1.0ex}
    \centering
    \includegraphics[width=1.0\linewidth]{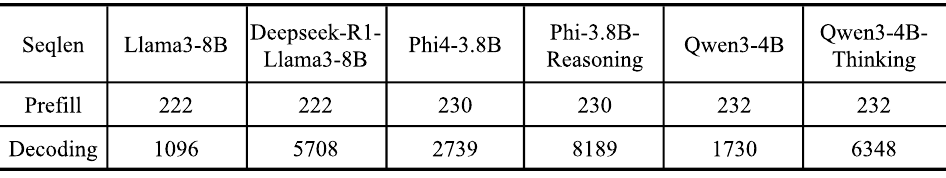}
    \label{table_2_1} 
    \vspace{-1.5em}
\end{table}

\subsection{Significant Accuracy Drop from Lossy Compression in Modern LLMs with Reasoning Capability}
\label{section_3_2}


In edge LLM inference scenarios with limited batching, lossy compression has been widely adopted as a primary acceleration technique. While effective in improving performance, lossy compression inevitably introduces accuracy degradation, and most existing approaches focus on minimizing this loss. However, techniques that are reported to incur negligible degradation on standard benchmarks often fail to generalize to more challenging tasks, where the resulting accuracy drop becomes substantial.

Table~\ref{6_2_accuracy_table} in section \ref{section_6_2} presents the measured accuracy after applying several state-of-the-art lossy compression techniques to various reasoning LLMs. As shown in the table, even lossy compressions that exhibit little performance degradation on relatively easy benchmarks suffer non-negligible performance drops on complex reasoning benchmarks.

This observation does not invalidate the overall effectiveness of lossy compression. Techniques such as quantization and pruning continue to provide significant performance gains while maintaining acceptable accuracy for less demanding workloads. However, considering that most recent small LLMs possess reasoning capabilities, and in contrast, quantization and pruning research is increasingly proposing aggressive techniques, relying solely on lossy compression going forward is not appropriate. To fully harness the performance of state-of-the-art LLMs in an edge environment, a method to accelerate low-batch LLMs while maintaining accuracy is absolutely essential.

\subsection{Potentials of Speculative Decoding and Limitations of Current Methods in Edge LLM Inference Environment}
\label{section_3_3}

Given the absence of a definitive solution for optimizing edge LLM inference, speculative decoding emerges as a promising approach for accelerating LLMs on consumer-grade xPUs. However, existing research on speculative decoding has primarily targeted large-scale models deployed in server environments, leaving several challenges unresolved in edge settings.

The most critical limitation is that many speculative decoding approaches rely on additional training. Although recent methods have reduced the training cost, such requirements remain impractical for users with resource-constrained devices. For instance, Eagle-3 speculative decoding model~\cite{eagle3}, a state-of-the-art training-based approach, requires approximately two days of training on four NVIDIA A100 GPUs to construct the draft model. This level of computational demand presents a significant barrier for edge users, who typically operate with limited hardware resources.

\begin{table}[t]
    \caption{Comparison of different speculative decoding methods}
    \vspace{-1.0ex}
    \centering
    \includegraphics[width=1.0\linewidth]{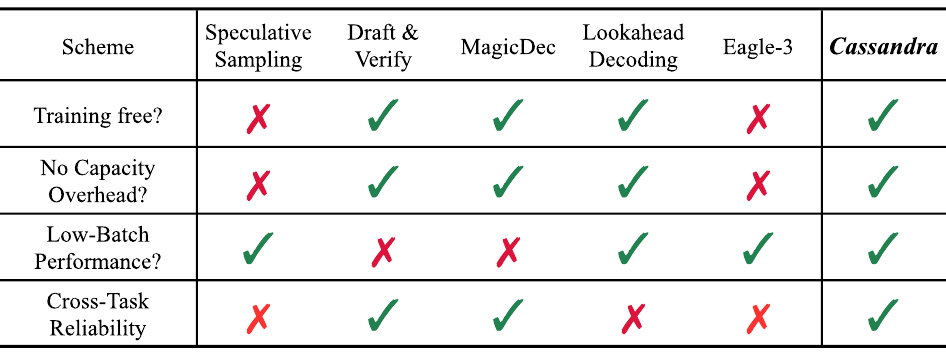}
    \label{table_3_3_speculative decoding} 
    \vspace{-3.0ex}
\end{table}

To address these issues, recent studies have explored training-free self-speculative decoding~\cite{Draft&Verify, QuantSpec, MagicDec}. These approaches construct the draft model directly from the original model, eliminating the need for additional training. For example, Draft\&Verify~\cite{Draft&Verify} generates draft outputs via layer skipping, while MagicDec~\cite{MagicDec} leverages sparse retrieval of the KV cache. Despite their advantages, existing self-speculative methods typically rely on coarse-grained approximations and primarily focus on reducing overhead in the attention mechanism and KV cache. As a result, although they demonstrate reasonable performance in batched inference scenarios, their benefits are often limited in low-batch, small-model settings that are common in edge environments.

In addition, several speculative decoding methods~\cite{speculativesampling, eagle3} employ independent draft models with separate weights and KV caches. This design introduces non-negligible memory overhead, which is particularly problematic for resource-constrained devices, as it reduces the maximum feasible sequence length under a fixed memory budget.

As summarized in Table~\ref{table_3_3_speculative decoding}, existing speculative decoding approaches exhibit distinct trade-offs, and no single method fully satisfies the requirements of edge deployment. Therefore, we propose Cassandra, a novel approach designed to overcome these limitations and enable efficient speculative decoding in edge LLM inference scenarios.

%% file: Contents/4_Speculative_Decoding_Algorithm.tex
\section{Cassandra Speculative Decoding Algorithm}
\label{section_4}

\subsection{Overview of Cassandra Algorithm}
\label{section_4_1}
To address the limitations of conventional speculative decoding in edge environments, Cassandra is designed around two key objectives: resource efficiency and low-batch performance.

First, Cassandra targets reliable operation under constrained compute and memory budgets. Approaches that require additional training are impractical for widespread deployment on edge devices, and maintaining a separate draft model introduces non-negligible memory capacity overhead. To overcome these issues, Cassandra adopts a training-free self-speculative decoding framework, eliminating the need for additional training while avoiding duplication of model parameters.

Second, Cassandra is optimized for performance in low-batch inference scenarios. Prior self-speculative decoding methods primarily focus on optimizing the attention layer and KV cache. Also, for computational efficiency on GPU, they usually adopt coarse-grained methods such as structured pruning or layer skipping to generate the draft model. However, such approaches are not sufficient in low-batch regimes. When the batch size is small and sequence lengths are moderate, the dominant bottleneck in the decode stage shifts from attention to the weight loading of the feed-forward network (FFN) layers. Therefore, improving performance in this setting requires not only optimizing attention-related operations but also effectively reducing the memory footprint of FFN weights. Furthermore, since the decode stage is typically memory-bound, minimizing data movement is more critical than improving raw computational efficiency.

\begin{figure}[t!]
\centering
\includegraphics[width=\colimgwidth]{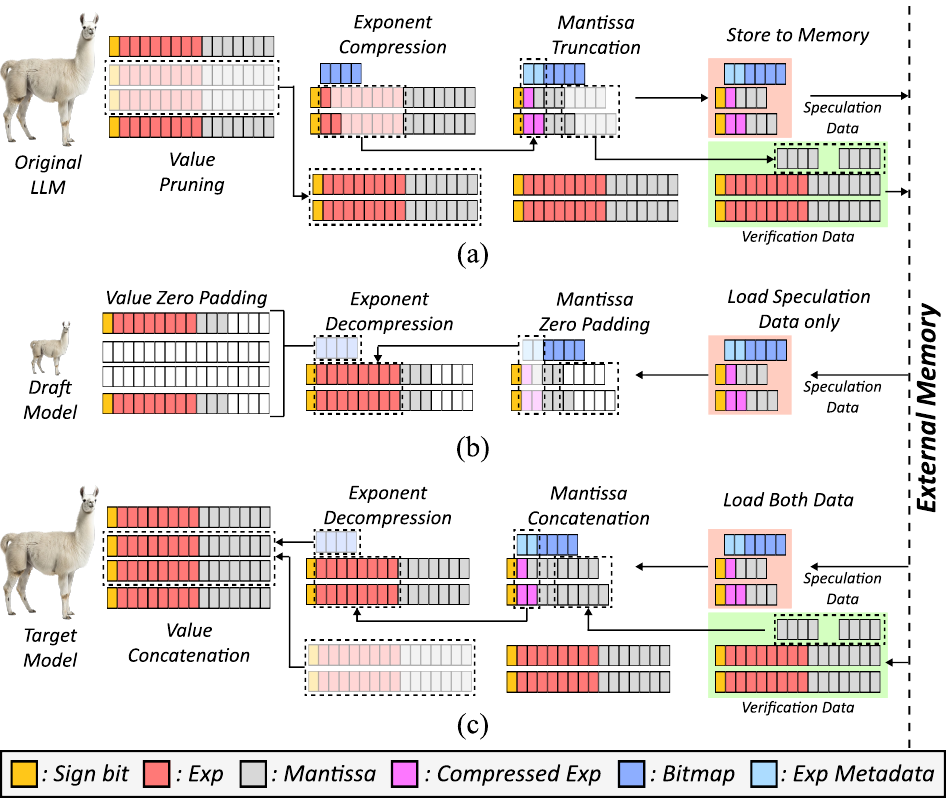}
\caption{Visualization of Cassandra Algorithm. (a) Cassandra's initial format transformation flow (b) Draft model inference (c) Target model inference}
\label{fig_4_1}
\end{figure}

Figure~\ref{fig_4_1} illustrates the core design of Cassandra. The weights and KV cache of the original model are first transformed into a specialized format and partitioned into two components: \textbf{speculation data} and \textbf{verification data}.

During draft inference, only the speculation data is loaded and reconstructed into the original format using zero-padding. Although the resulting draft model is executed using standard floating-point units and does not reduce arithmetic complexity, it significantly lowers memory bandwidth requirements, enabling faster execution.

During target model inference, both the speculation and verification data are loaded and fully reconstructed to recover the original model representation. Unlike conventional speculative decoding methods~\cite{speculativesampling, eagle3}, Cassandra does not require an independent draft model. Instead, the draft model operates on a strict subset of the original model parameters, allowing speculative decoding to be performed without additional memory capacity overhead.

\subsection{Step-by-Step Format Transformation Flow in Cassandra}
\label{section_4_2}
In this section, we elaborate on how the original model undergoes a format transformation process to be partitioned into speculation data and verification data. In the subsequent paragraphs, the data selected through the pruning or truncation process constitute the speculation data. Unlike lossy compression, the data that are not selected during this process are not discarded; instead, they become the verification data and are utilized for target model inference.

~\textbf{Unstructured Value Pruning.} First, Cassandra performs unstructured pruning on the weights and KV caches of the original model. Regarding weights, we employ the activation-aware weight pruning technique proposed in Wanda~\cite{Wanda}. Leveraging the observation that activations exhibit large magnitudes in specific channels regardless of the input, Wanda computes the L2 norm of activations using calibration data and multiplies the resulting values element-wise by the weights to determine importance scores for pruning.

In contrast to weights, Cassandra applies per-token magnitude-based pruning for the KV cache. According to Mustafar~\cite{Mustafar}, this approach is highly effective for Key caches due to the presence of prominent channel-wise outliers. Although Value caches do not exhibit a similarly distinct outlier distribution, per-token pruning remains effective, as it is functionally equivalent to output-aware pruning within the attention mechanism.

Since these methods are designed primarily to preserve the output of each layer, they are more advantageous in terms of acceptance rate relative to compression ratio compared to structured pruning, which utilizes pre-determined masks for computational efficiency. Also, both methods involve minimal to negligible calibration costs, aligning well with Cassandra's design philosophy of being training-free.

~\textbf{Mantissa Truncation.} In addition to pruning, we further reduce the representation cost of each value by truncating mantissa bits. While quantization is the conventional approach for reducing bit-width, we instead adopt a naive mantissa truncation scheme. This choice significantly lowers the overhead of format reconstruction compared to quantization. Furthermore, unlike quantization, which alters the numerical representation, mantissa truncation preserves a subset of the original bits. As a result, the draft model can be interpreted as a strict subset of the target model, which is a key property that enables Cassandra to avoid additional memory capacity overhead.


\textbf{Exponent Compression.} The application of the two aforementioned algorithms still results in limited compression ratios. This limitation stems from the uncompressed exponent, which accounts for 50\% of the bit-width in the standard BFloat16 datatype. Therefore, to achieve significant performance improvements, additional compression must be applied.

\begin{figure}[t]
\centering
\includegraphics[width=\colimgwidth]{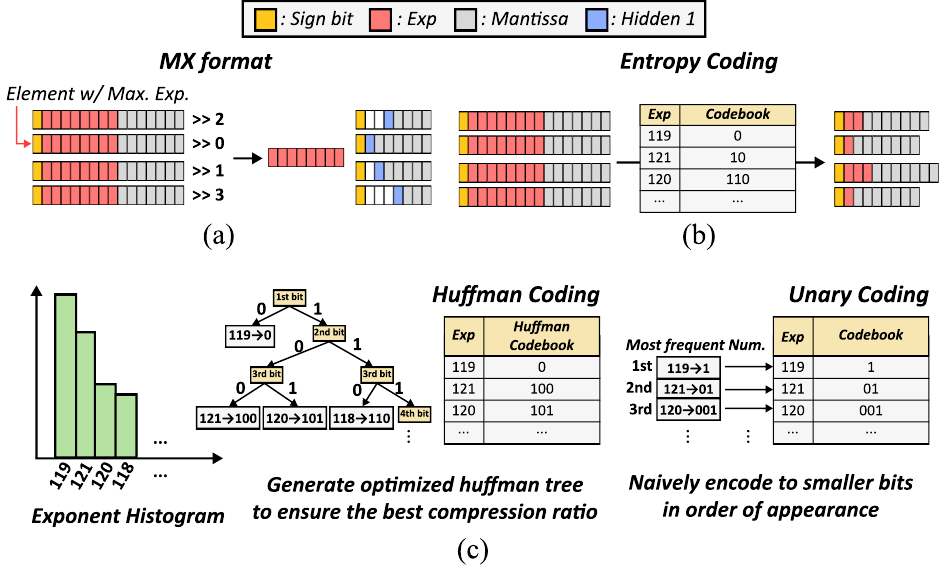}
\caption{(a) MX format, (b) Entropy Coding, (c) Huffman and Unary Coding}
\label{fig_4_2}
\end{figure}

One possible approach is to adopt the MX format~\cite{microscaling}, in which multiple floating-point values share a common exponent. This format has been widely utilized in deep learning training and inference due to its efficiency. However, converting a model trained in a standard floating-point format into MX format inevitably introduces some degree of accuracy degradation. Although this degradation is often small, we argue that a complementary approach with no risk of accuracy loss is also desirable. From this perspective, we identify entropy coding as a suitable method for lossless exponent compression.

\begin{figure}[t]
\centering
\includegraphics[width=\colimgwidth]{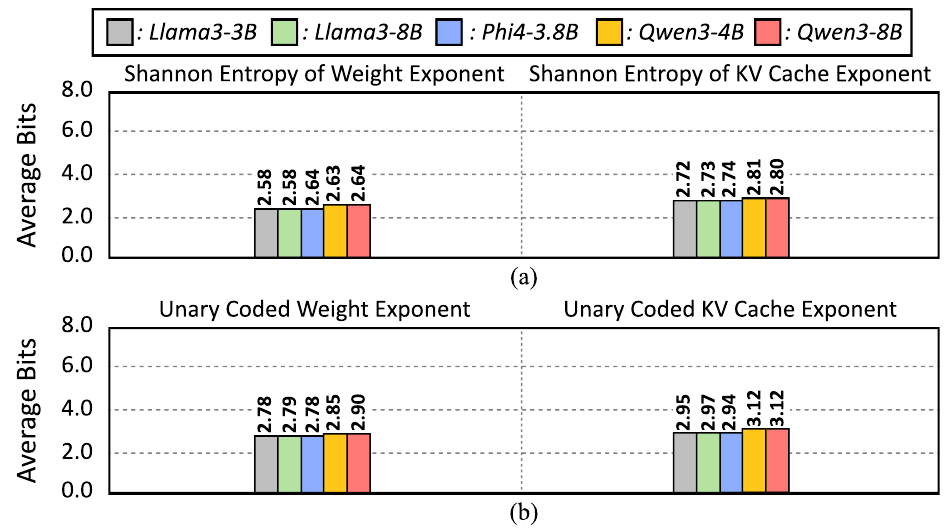}
\caption{(a) Average Shannon entropy of exponent in weight and KV cache. (b) Average exponent bits of unary-coded weights and KV cache}
\label{fig_4_3}
\end{figure}

Entropy coding is a lossless compression technique that assigns variable-length codes based on the frequency distribution of values. Shannon entropy~\cite{ShannonEntropy} provides a theoretical lower bound on the average number of bits required to represent such data. Previous work~\cite{dfloat11} has shown that the exponent values of BFloat16-trained weights exhibit a Shannon entropy of approximately 2.6 bits. As illustrated in Figure~\ref{fig_4_3}(a), we further observe that the exponent distribution of the KV cache also has low entropy, averaging around 2.7 bits. These results suggest that, by applying appropriate entropy coding to both weights and KV cache, it is theoretically possible to achieve a lossless reduction of more than 5 bits.


Among entropy coding techniques, Huffman coding~\cite{HuffmanCoding} is one of the most widely used methods, and several prior studies~\cite{dfloat11, Huff-LLM} have explored its application to LLM compression. However, implementing Huffman decoding efficiently on xPUs presents significant challenges. A conventional LUT-based decoding approach~\cite{Huffman-GPU-LUT} requires a lookup table with $2^N$ entries, where $N$ denotes the number of unique symbols. In LLMs, where $N$ can be as large as 32, such an approach becomes impractical. While hierarchical codebooks~\cite{dfloat11} can mitigate this issue, they introduce additional decoding complexity and latency. As a result, the overhead of decoding may outweigh the benefits of compression, leading to worse overall system performance compared to the BFloat16 baseline.


To overcome this issue, we propose a much simpler method: Unary coding can be a solution for low-overhead lossless exponent compression. Unary coding is a method of assigning codes simply by increasing the length by one, corresponding to the frequency of a particular element. In Cassandra, unary coding is implemented by encoding numbers in a format where $N$ consecutive zeros are followed by a final one, such as $1, 01, 001,$ and so on. Exponents that appear more frequently are assigned fewer bits. The key advantage of this method lies in its explicit boundary representation: every codeword is terminated by a \textbf{1}, allowing unambiguous identification of code boundaries directly from the compressed bitstream. As a result, unlike conventional entropy coding schemes, which often require sequential parsing or LUT-based decoding, unary coding can be implemented using simple, fully parallel digital logic.


As shown in Figure~\ref{fig_4_3}(b), we confirmed that unary coding can achieve an average exponent compression of 2.85 bits. Although its compression efficiency is slightly lower than that of Huffman coding, it still provides a substantial improvement over the original floating-point representation. Also, from the perspective of overall system performance, the reduction in decoding overhead achieved by using unary coding provides a more positive impact than the slight loss in compression ratio.

Unary coding preserves full numerical accuracy, whereas the MX format offers higher compression efficiency at the cost of slight accuracy degradation. Consequently, neither approach can be considered universally superior. To accommodate this trade-off, Cassandra supports both methods. We refer to the unary coding-based, lossless configuration as Cassandra-1, and the MX-based, higher-performance configuration with minor accuracy loss as Cassandra-2. This design enables users to select the most appropriate configuration based on their accuracy and performance requirements. Although mantissa truncation was introduced earlier for clarity, exponent compression is applied prior to mantissa truncation in the actual algorithm, as illustrated in Figure~\ref{fig_4_1}.

\subsection{Design Space Exploration of Cassandra Algorithm: Trade off between Acceptance Rate and Compression Ratio}
\label{section_4_3}
In LLM inference with lossy compression, a higher compression ratio typically results in a larger accuracy drop. Thus, the acceptable accuracy drop determines the final compression ratio. For this reason, combining multiple lossy compression algorithms is often avoided in standard LLM inference scenarios.

However, speculative decoding uses the target model for verification, essentially eliminating the accuracy drop, regardless of the draft model's compression ratio. Instead, in this case, the compression ratio and the acceptance rate are in a trade-off relationship. At this point, the draft model does not need to generate the correct complete sentence by itself; it only needs to retain the ability to predict a few tokens correctly. In this case, the conditions for generating the optimal draft model differ from the methods used in lossy compression.

\begin{figure}[t]
\centering
\includegraphics[width=\colimgwidth]{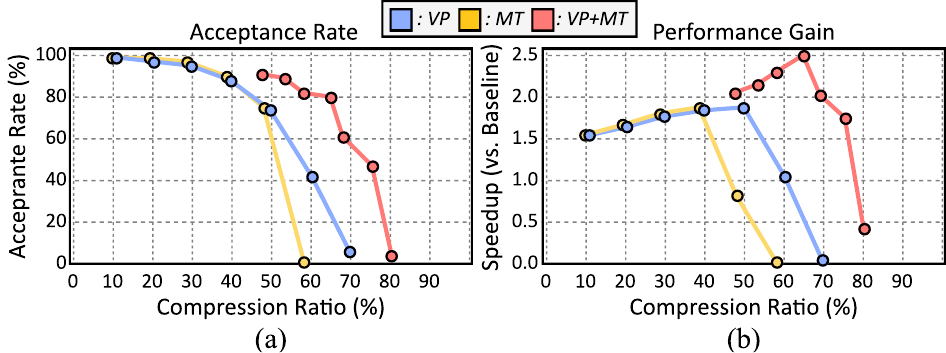}
\caption{(a) Acceptance rate according to compression ratio($\gamma$=5). (b) Ideal performance improvement compared to baseline. VP means value pruning, MT means mantissa truncation, VP+MT means applying both schemes.}
\label{fig_4_4}
\vspace{-1.0em}
\end{figure}

Figure~\ref{fig_4_4} shows the acceptance rate versus compression ratio when using the Deepseek-R1-Distillated-Llama-8B model, with the draft model created using value pruning and mantissa truncation separately, and when using both schemes together. At this point, using value pruning and mantissa truncation together demonstrates a robust acceptance rate compared to using either method alone. This result demonstrates that Cassandra's format transformation method, which integrates multiple schemes, is suitable for self-speculative decoding.

%% file: Contents/5_Hardware_Architecture.tex
\section{Cassandra Hardware Architecture}
\label{section_5}

\subsection{Cassandra Encoder $\&$ Decoder Architecture}
\label{section_5_1}
To actually enhance the performance of xPU with Cassandra, low-overhead decompression is necessary. In contrast, operations such as MX format to floating point transformation, unary coding, and bitmap-based de-sparsification all require bit-level decoding computations. Performing these operations sequentially on a typical SIMD core cannot fully leverage hardware parallelism. Also, utilizing Cassandra necessitates online KV cache format conversion, and the operations used in this process, such as top-$k$ computations, can also be a potential burden. Therefore, we propose a Cassandra decoder and encoder to achieve high performance and low overhead format transformation, enabling the efficient use of Cassandra across various computing units.


\begin{figure}[t]
\centering
\includegraphics[width=\colimgwidth]{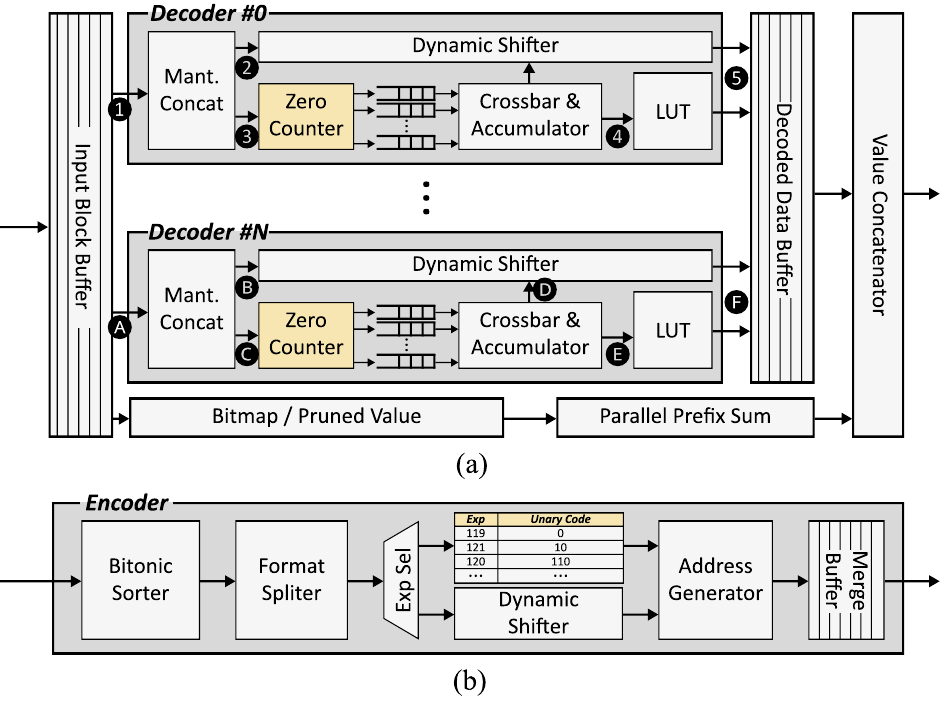}
\caption{(a) Microarchitecture and dataflow of Cassandra decoder. (b) Microarchitecture of Cassandra encoder.}
\label{fig_5_2}
\end{figure}
\begin{figure}[t]
\centering
\includegraphics[width=\colimgwidth]{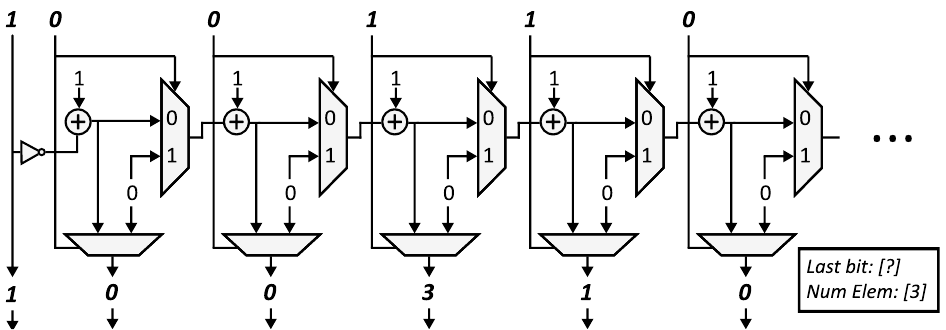}
\caption{Microarchitecture of parallel zero counter.}
\label{fig_5_1}
\vspace{-1.5em}
\end{figure}

Figure~\ref{fig_5_2} illustrates the architecture of our decoder and encoder, and how these hardware components operate across different dataflows. Decoder \#0 in Figure~\ref{fig_5_2}(a) demonstrates the decoding dataflow when using the unary coding for exponent compression. Initially, the low bits and high bits of the truncated mantissa are concatenated in the \myfilledcircled{1}~\textbf{mantissa concatenator}. Subsequently, the sign and mantissa are fed to the \myfilledcircled{2}~\textbf{dynamic shifter} and await exponent decoding. The exponent is uniformly sliced into 8-bit chunks and sent to the \myfilledcircled{3}~\textbf{parallel zero counter}.

Figure~\ref{fig_5_1} presents the microarchitecture of a parallel zero counter. For each bit position, the module outputs the cumulative number of preceding zeros up to that position. In parallel, it also produces the total count of input ones and the value of the final bit. The output from the parallel zero counter passes through a zero eliminator and enters a queue, where it is used as an index for the \myfilledcircled{4}~\textbf{LUT} for unary decoding. 


Although we arbitrarily segmented the sequentially encoded exponents into 8-bit chunks and input them into a parallel zero counter, obtaining the correct decoding value necessitates information from the preceding bits. This challenge can be addressed by recalculating in each chunk, which involves concurrently receiving the last bit of the preceding chunk and the count of consecutive zeros last obtained from that preceding chunk. Algorithm~\ref{algorithm_1} describes the entire process of exponent decoding within a parallel zero counter. After these processes, a \myfilledcircled{5}~\textbf{bitmap-based} de-sparsification operation is performed to derive the final output.

\begin{algorithm}[t] 
\caption{Pseudo Code of Parallel Unary Decoding\label{alg:search}}
\label{algorithm_1}
\KwIn{N-bit unary coded exponent $U$, the number of element $numel$}
\KwOut{Set of decoded 8bit exponents $Exp$}

\tcc{Parallel Zero Counting}

$[Chunk_0, Chunk_1, ..., Chunk_{[\frac{N}{8}]}] \leftarrow \func{DevideInto8bitChunk}(U)$

$i,k,l, sum, idx \leftarrow 0$

\While{$i < [N//8] $}{
    $cnt\leftarrow 0$,
    $reorganized_i \leftarrow Chunk_i[7]$
    
    $num\_ones_i \leftarrow \func{NumberofOnes}(Chunk_i)$
    
    \For{$j$ in $range(8)$ } {
        \eIf{$Chunk_i[j]==0$}
        {
            $cnt \leftarrow cnt + 1$
        }{
            $output_{i}.append(cnt)$, \
            $cnt\leftarrow 0$
        }
    }
    $i \leftarrow i+1$
}

\tcc{Parallel Reorganization}

\While{$k < [N//8] $}{
    \eIf{$reorganized_k==0$}
    {
        $sum \leftarrow sum + cnt_k$
    }{
        $output_{k}[0]\leftarrow output_{k}[0] + sum$ \\
        $sum \leftarrow 0$
    }
    $k \leftarrow k+1$
}

\While{$l < [N//8] $}{
    \For{$m \; in\;  range(num\_ones_{l}) $}{
    $Exp_{idx} \leftarrow \func{UnaryCodebook}(output_{l}[m])$
    $idx \leftarrow idx+1$ \\
    \If{$idx > numel$}{
    \KwRet{$Exp$}
    }
    }
}

\KwRet{$Exp$} 

\end{algorithm}
Decoder \#N in Figure~\ref{fig_5_2}(a) demonstrates the decoding dataflow when using the MX format. The low-bits and high-bits of the mantissa are combined by the \myfilledcircled{A}~\textbf{mantissa concatenator} first. However, in this case, after the mantissa is combined, the mantissa is fed into the input of both \myfilledcircled{B}~\textbf{dynamic shifter} and \myfilledcircled{C}~\textbf{parallel zero counter}. In this case, the exponent is first passed through a crossbar and broadcast to the accumulator corresponding to each element. Subsequently, the parallel zero counter counts the number of zeros in the mantissa, determining how many bits each mantissa should be shifted and how much should be subtracted from the exponent. Since it is guaranteed that one counter processes one mantissa, adjustment with adjacent chunks is not necessary. Once the shifted value for each mantissa is determined, this value is sent to the \myfilledcircled{D}~\textbf{dynamic shifter} to perform mantissa shifting and simultaneously subtracted from the exponent in the \myfilledcircled{E}~\textbf{accumulator}. Afterwards, the decoding process concludes by performing \myfilledcircled{F}~\textbf{bitmap-based} de-sparification, similar to Cassandra-1.


The operation of the Cassandra encoder is comparatively simpler. When data is first input, the encoder finds the top-$k$ values in this data and divides them into a speculation group and a verification group. The verification group is stored in a buffer immediately, along with a bitmap, without any further transformation. The data belonging to the speculation group undergoes different exponent compression processes depending on whether they use MX format or unary coding, followed by mantissa truncation, and then they are stored in the buffer. Subsequently, the values stored in the buffer are saved to main memory. Since weights can be formatted offline before storage, this encoder is primarily used for online formatting of the KV cache.

\subsection{Integrating Cassandra with xPU}
\label{section_5_2}
\begin{figure*}[t!]
\centering
\includegraphics[width=\fullimgwidth]{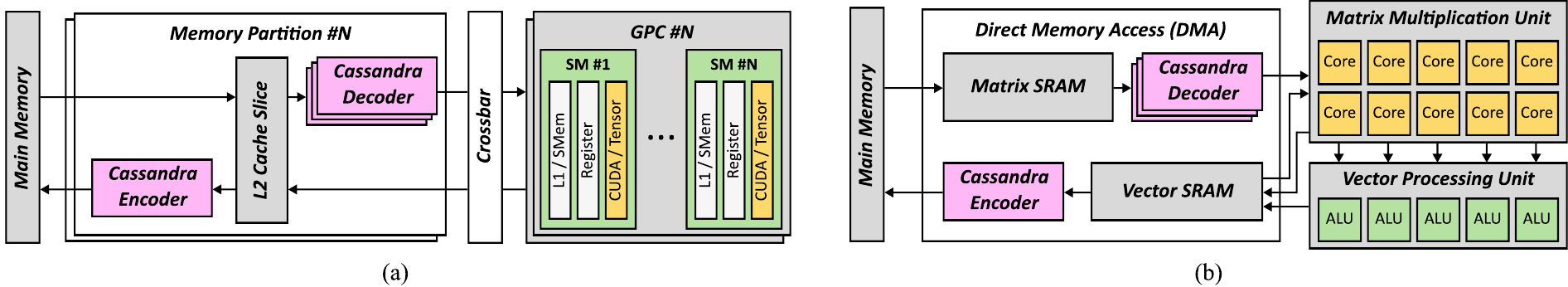}
\caption{Overall architecture of (a) Cassandra-integrated GPU and (b) Cassandra-integrated systolic array based NPU}
\label{fig_5_3}
\end{figure*}

Cassandra is designed to operate alongside the memory system of a GPU and NPU. A GPU utilizes a partitioned memory system architecture, where each memory channel has its own dedicated L2 cache slice and memory controller. Furthermore, all data originating from the main memory must pass through the L2 cache before being delivered to the L1 cache or shared memory. Figure~\ref{fig_5_3}(a) illustrates how Cassandra can be integrated into such a system. Because sharing data between memory partitions is challenging in a GPU, Cassandra encoder and decoder should be installed independently for each memory partition.
In this configuration, the decoder is placed between the L2 cache and the interconnect, and the encoder is situated between the main memory and the L2 cache. This placement allows for the efficient utilization of both memory bandwidth and L2 cache capacity and bandwidth. In this case, Cassandra's encoder and decoder are managed by the memory controller, similar to the L2 cache.

To distinguish standard data types (e.g., activations) from specially formatted weights and KV cache, the user must explicitly specify the data type being stored (i.e., floating-point or Cassandra format) during the memory allocation process. Data of the standard datatype and the Cassandra datatype are then stored in separate virtual pages. To distinguish the data type held by each page, the page table and TLBs must be provisioned with spare bits. Since commercial GPUs from vendors like Nvidia and AMD already include spare bits in their page table entries, this modification introduces no additional overhead.~\cite{ecco}.

For an NPU with a DMA-based memory system, as shown in Figure~\ref{fig_5_3}(b), it is reasonable to place both the encoder and decoder within the DMA. This structure is similar to the one in GPUs, but in this case, the encoder and decoder should be managed by the DMA controller. Also, unlike GPUs, many NPUs do not use a virtual memory system. Therefore, in this scenario, the physical memory addresses for storing the standard datatypes and Cassandra datatypes must be separated, and this information must be pre-stored in the DMA controller to ensure appropriate encoding and decoding.

Decoding overhead in Cassandra can slow down the read speed of weights and the KV cache, degrading overall system performance, while the encoder's performance is irrelevant to this. Hence, the decoder should be sufficiently added to match the maximum throughput of the L2 cache, while a comparatively smaller number of encoders may be sufficient.

\subsection{Superblock based Data Management}
Variable sequence length in unary encoding, as well as the potential difference in the ratio between speculation data and verification data that may arise when sparsity changes, makes memory mapping more challenging. Specifically, in a cache-based memory system like a GPU, this difficulty in mapping data to perfectly fit a cache block can lead to performance degradation due to redundant data loads. To eliminate this potential performance bottleneck, we propose a superblock-based memory mapping scheme. For a cache-based system, Cassandra sets a group of multiple cache blocks, which we refer to as a superblock, as the fundamental unit for load and eviction. This superblock is fully packed with cache blocks containing various data types, such as bitmaps, variable exponents, and mantissas. Initially, when data is loaded, the entire superblock is loaded at once and stored in the L2 cache. These data are subsequently fed into the decoder.

\label{section_5_3}
\begin{figure}[t]
\centering
\includegraphics[width=\colimgwidth]{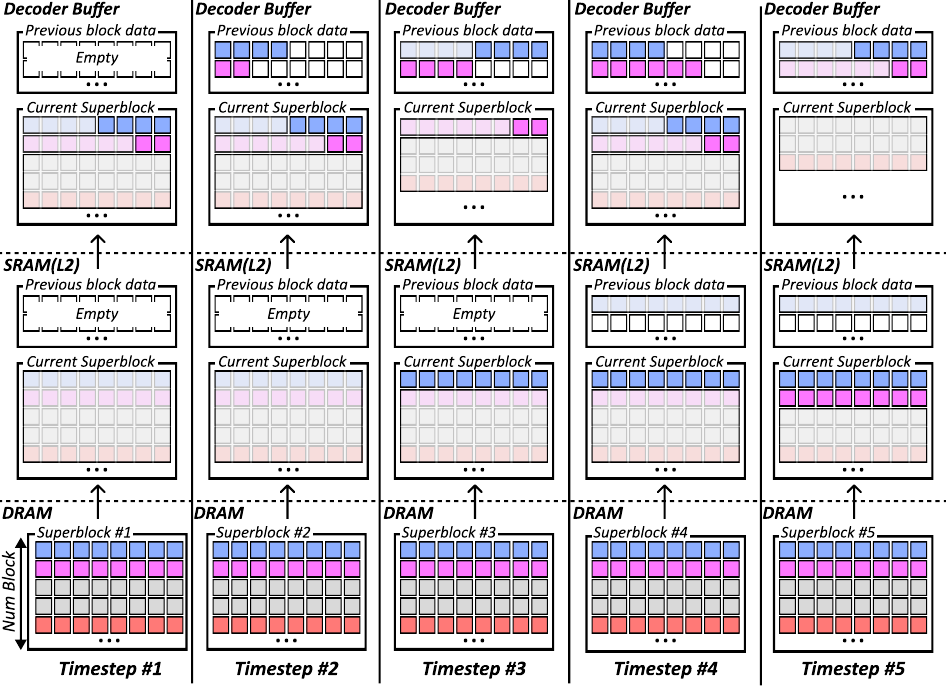}
\caption{Visualization of superblock-based data management.}
\label{fig_5_4}
\vspace{-1.0em}
\end{figure}

During the decoding process, depending on the sparse ratio or the length of the exponent, some data types might not fully utilize all the data contained within a single cache block, resulting in leftover data. The decoder stores this leftover data in its internal buffer and concatenates it with the data from the next cache block of that type when it is read.

If the size of the data in the decoder buffer exceeds 128 bytes, the memory controller skips sending the corresponding type of cache block to the decoder. It also keeps track of how many times the block of that type has been skipped for the next read. In this manner, the memory controller manages the address values of each type of cache block that should be fed into the decoder next, sending them to the decoder as needed.

This approach guarantees contiguous data reads from global memory and dense memory mapping, preventing potential degradation in memory performance. This method also works efficiently in scratchpad-based devices, such as NPUs, where issues like row buffer misses and page misses can still occur. For an NPU, the same mechanism can be applied by directly setting an arbitrary block size, rather than a cache block size.

%% file: Contents/6_Methodology_and_Evaluation.tex
\section{Evaluation}
\label{section_6}


\begin{figure*}[t]
\centering
\footnotesize
\includegraphics[width=\fullimgwidth]{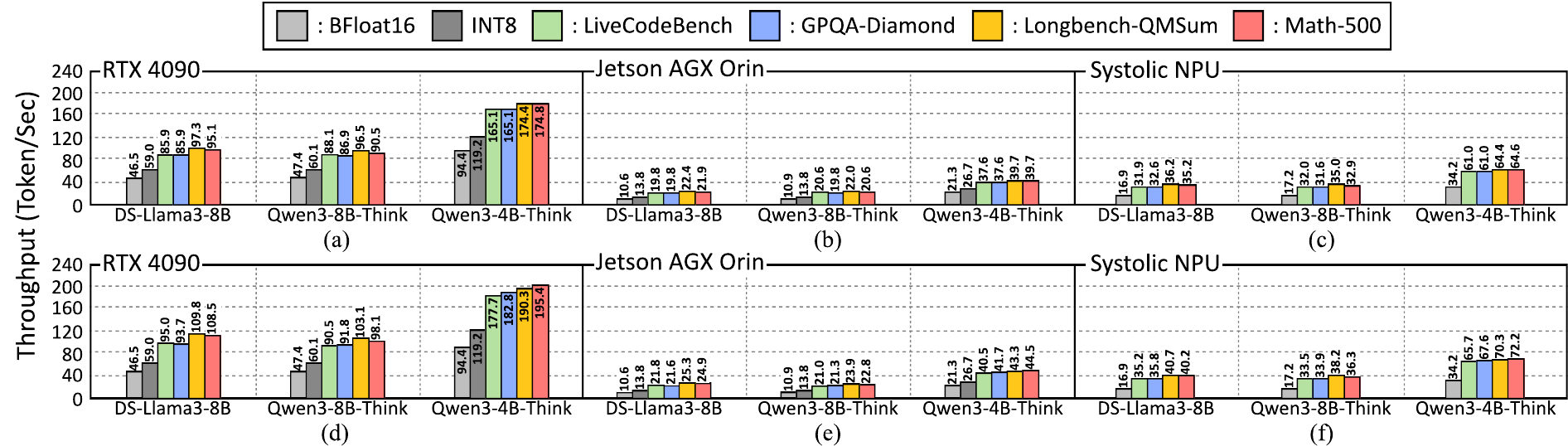}
\caption{Normalized performance gain through Cassandra on various hardware \& benchmark. (a) RTX 4090 + Cassandra-1, (b) Jetson AGX Orin + Cassandra-1, (c) Systolic Array NPU + Cassandra-1, (d) RTX 4090 + Cassandra-2, (e) Jetson AGX Orin + Cassandra-2, (f) Systolic Array NPU + Cassandra-2}
\label{fig_6_1} 
\end{figure*}

\subsection{Implementation of Cassandra}
\label{section_6_1}
~\textbf{Cassandra Software Implementation.} To measure the accuracy of Cassandra, we developed a software emulator utilizing PyTorch and custom CUDA kernels. Although this software emulator cannot achieve performance gain on a GPU, it accurately models the pruning, truncation, unary coding, and MX format conversion that occur during the format transformation process in Cassandra.

~\textbf{Cassandra Hardware Implementation.} The performance of Cassandra cannot be characterized solely by hardware cycles; it also depends on the acceptance rate. Therefore, to measure the performance gain obtained by Cassandra, we first determine four different scenarios from benchmarks~\cite{LiveCodeBench, PRM800K(source_of_math_500), Longbench, GPQA-Diamond} and apply them to Cassandra with 40\% pruned and 4-bit truncated weights and 4-bit truncated KV cache. In addition, we measure the average input and output context lengths, as well as the acceptance rate for those scenarios. Finally, we use this information with the cycle information from our hardware simulator to quantify overall system performance.

To measure the performance of Cassandra when integrated with a GPU, we implement the Cassandra encoder and decoder on Accel-Sim~\cite{Accel-Sim}. We use Nvidia RTX 4090~\cite{RTX4090} and Nvidia Jetson AGX Orin~\cite{Jetson} as GPU baselines. Since Accel-Sim currently does not support the Ada-Lovelace architecture used in the RTX 4090, we use traces from an Ampere architecture GPU and set the configuration similar to RTX 4090.

Additionally, to evaluate the performance of Cassandra-integrated NPU, we implemented a cycle-level simulator by extending Scale-Sim~\cite{Scale-Sim} and the LPU simulator~\cite {LPU, oaken}. Referring to the specifications of commercial consumer-grade GPUs, we designed this NPU to feature a 64 TFLOPS MAC unit and a memory bandwidth of 273 GB/s, utilizing 128GB of LPDDR5X memory.

To analyze the power and area overhead of Cassandra, we first implement an end-to-end Cassandra decoder and encoder with SystemVerilog. We also implement a simple consumer-grade NPU, consisting of a systolic array, VPU, DMA, and scratchpad, and integrate it with Cassandra. Both Cassandra and NPU are synthesized using the Synopsys Design Compiler with a 28nm technology node, and the area of the SRAM used was obtained by utilizing the Samsung 28nm SRAM Compiler.

\begin{table}[t]
\caption{Zero-shot accuracy results on Reasoning Benchmarks}
\centering
\includegraphics[width=\colimgwidth]{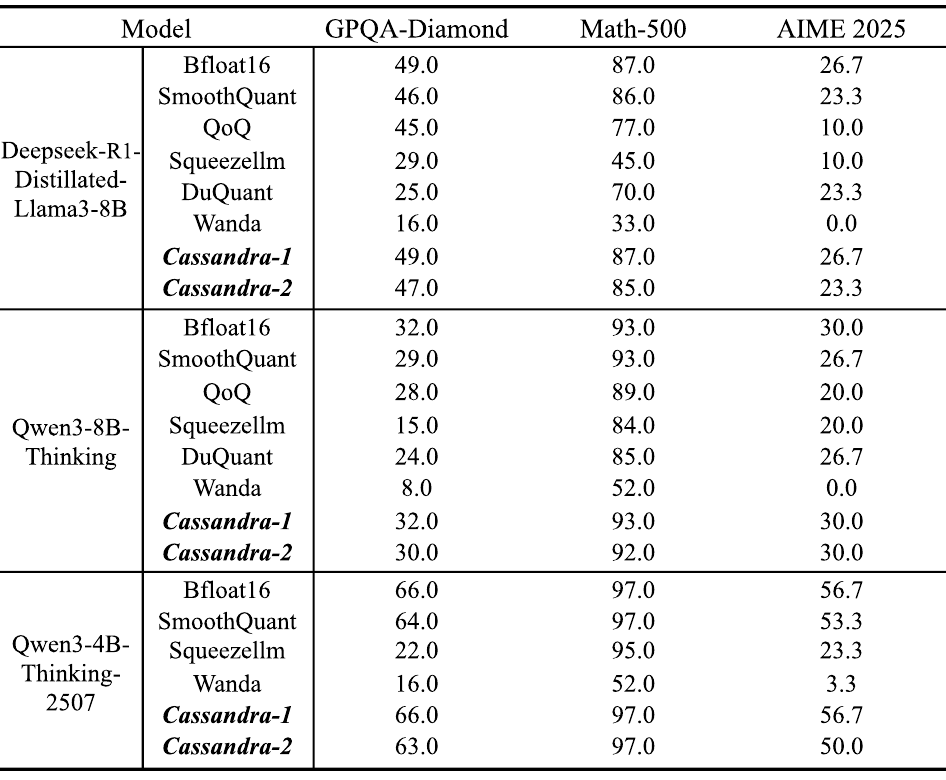}
\label{6_2_accuracy_table}
\vspace{-3.0em}
\end{table}

\subsection{Zero-Shot Accuracy Evaluation}
\label{section_6_2}

In this part, we compare state-of-the-art lossy compression methods~\cite{Smoothquant, qserve, duquant, Wanda, squeezellm} with Cassandra across various benchmarks. The pruning ratio and precision used for lossy compression are identical to those in the original papers. We use Deepseek-R1-Distillated-Llama3-8B~\cite{Deepseek-R1-Distillated-Llama3-8B}, Qwen3-8B-Thinking~\cite{Qwen3-8B}, and Qwen3-4B-Thinking-2507~\cite{Qwen3-4B-Thinking-2507} as LLM models and use AIME2025~\cite{AIME2025}, Math-500~\cite{PRM800K(source_of_math_500)}, GPQA-Diamond~\cite{GPQA-Diamond} as benchmarks. Due to the limited GPU resources, we randomly sampled 100 questions from the GPQA-Diamond and Math-500, while the AIME2025 benchmark was tested using all its questions. Table~\ref{6_2_accuracy_table} compares Cassandra with other lossy compression techniques with these models and benchmarks. 

As illustrated in the table, all lossy compression methods experience a certain degree of accuracy degradation. Smoothquant, which utilizes relatively higher bitwidth, exhibits only a marginal drop in accuracy compared to the BFloat16 baseline; however, its compression ratio remains modest at approximately 50\%.
In contrast, all other lossy compression techniques suffered significant accuracy losses across reasoning benchmarks. Notably, when applied to DeepSeek-R1-Distillated-Llama3-8B, Wanda failed to produce a single correct answer on the AIME 2025 benchmark.
Conversely, Cassandra-1 maintained accuracy levels identical to BFloat16. Also, Cassandra-2, which prioritizes inference speed at the expense of some precision, demonstrated a robust accuracy profile comparable to that of Smoothquant.


As illustrated in the table, DuQuant~\cite{duquant} and QoQ~\cite{qserve} were not evaluated on the Qwen3-4B-Thinking-2507 model. These algorithms maintain model accuracy by employing Hadamard rotations to smooth out the distribution of values, thereby reducing the impact of outliers. However, since Hadamard matrices are restricted to specific discrete dimensions, their open-source implementations are not directly compatible with the unique hidden dimension of Qwen3-4B-Thinking-2507. Applying these methods would require additional pre-processing, such as zero padding.

\subsection{Performance}
\label{section_6_3}
Figure~\ref{fig_6_1} presents Cassandra's overall performance gain obtained from our simulator. Since Cassandra's performance varies depending on the model and the benchmark, we express its performance using the throughput gain achievable across four different scenarios, three models, and various exponent compression techniques. Furthermore, the length of the draft tokens $\gamma$ is set to the value that yielded optimal performance within the range of 3 to 5.
Furthermore, we included SmoothQuant-based W8A8 integer quantization~\cite{Smoothquant} as a GPU baseline, utilizing the official INT8 quantization implementation provided by vLLM~\cite{W8A8-INT8-Dynamic}. While we also measured the throughput of FP8 dynamic quantization~\cite{W8A8-FP8-Dynamic} on an RTX 4090, it consistently showed a slight performance deficit in the decode stage compared to INT8. However, this performance gap was marginal, staying under 3\% for all tested models.

\begin{table}[t]
\caption{Average Acceptance Rate on Various Models and Benchmarks.}
\centering
\includegraphics[width=\colimgwidth]{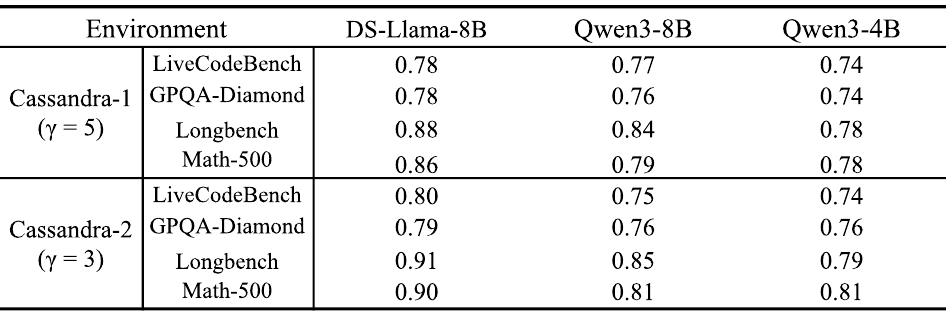}
\label{table_6_3}
\vspace{-0.20in}
\end{table}

As shown in the figure, Cassandra achieved a performance improvement of 1.78$\times$ to 2.41$\times$ compared to BFloat16 baseline. Two primary factors influence the performance of Cassandra. One factor is the disparity between models. As shown in Table~\ref{table_6_3}, under identical configurations and benchmarks, DeepSeek-R1-Distillated-Llama3-8B typically exhibited the highest acceptance rate, followed by Qwen3-8B and Qwen3-4B-Thinking-2507, respectively. The second factor is the difference of benchmark. Across all models, Cassandra performed slightly better on Longbench-QMSum and Math500 than on LiveCodebench and GPQA-Diamond. Nevertheless, Cassandra shows less performance variance between benchmarks compared to other methodologies such as Eagle-3~\cite{eagle3} and Lookahead Decoding~\cite{Lookahead-Decoding}; this is discussed in further detail in section~\ref{section_7_2}.




\subsection{Area and Power Analysis}
\label{section_6_4}

Table~\ref{table_6_4} shows the power and area overhead of Cassandra on an NPU with 64TFLOPS and 9MB scratchpad. The total bandwidth of the scratchpad is 1024 bytes, and 40 decoders were added to ensure that the scratchpad bandwidth does not become a bottleneck during the read process. As shown in the table, the Cassandra system only incurs an area overhead of around 2\% relative to the NPU.

Note that directly comparing the power and area of Cassandra with the RTX 4090 and Jetson AGX Orin, which are fabricated using a 5nm and 8nm node, respectively, is not appropriate. However, despite using highly advanced nodes, these devices consume very large areas ($609mm^2$ and $455mm^2$, respectively) and power ($450W$ and $200W$, respectively). Considering the small overhead in an area-efficient NPU, we believe that Cassandra will show even smaller area and power ratios on commercial GPUs.

\begin{table}[t]
\caption{Area and power overhead of Cassandra on 64TFLOPS NPU.}
\centering
\includegraphics[width=\colimgwidth]{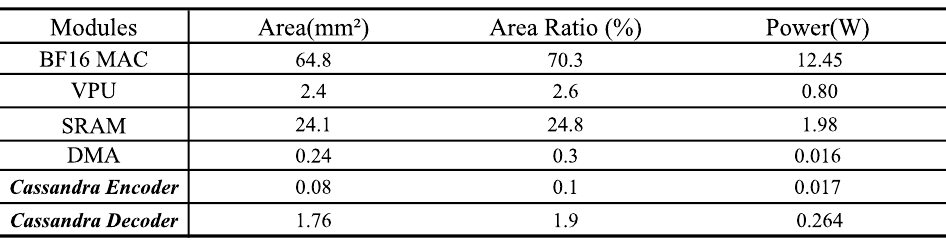}
\label{table_6_4}
\vspace{-0.20in}
\end{table}

%% file: Contents/7_Analysis_and_Discussion.tex
\section{Analysis and Discussion}
\label{section_7}


\subsection{Cassandra and Quantization: From Performance Rivalry to Algorithmic Synergy}
\label{section_7_1}
Quantization has emerged as a critical methodology for optimizing LLM inference performance. In particular, high-bit quantization formats, such as INT8, FP8 and MXFP8, are widely recognized for their robustness; this may raise questions regarding the necessity of lossless acceleration techniques like Cassandra. Nevertheless, Cassandra offers distinct advantages over the mere employment of high-bit quantization.

\subsubsection{Unpredictable Accuracy Degradation in Quantization}
According to previous studies~\cite{Quantization-Hurts-Reasoning, llama70b-quant, llama405b-quant}, even quantization techniques generally known to maintain accuracy can suffer from significant performance degradation under specific conditions. For instance, applying W8A8 SmoothQuant~\cite{Smoothquant} to Llama3.1-405B results in an average accuracy decline of 10.86\% across the OpenLLM Leaderboard-v2 datasets~\cite{llama405b-quant}. Such unpredictable degradations manifest across a wide range of model scales, from Qwen2.5-1.7B to Llama3.1-405B. Furthermore, quantization alters the model's probability distribution, potentially leading to unpredictable behaviors. For instance, a study~\cite{Accuracy-is-Not-All-You-Need} reported that applying W8A16 GPTQ~\cite{GPTQ} to Qwen2-1.5B~\cite{Qwen2-1.5B} resulted in only a 0.3\% accuracy drop on the GSM8K benchmark~\cite{GSM8K}; however, 6.37\% of the actual predicted answers changed compared to the original model.

For these reasons, despite the high efficiency of quantization, there remains a strong demand for lossless LLM inference acceleration, where speculative decoding is often employed independently without quantization. Furthermore, recent studies~\cite{dfloat11, LILO, Zipserv} have begun exploring the use of lossless compression techniques to accelerate LLM inference, without relying on speculative decoding.

\subsubsection{Performance Superiority of Cassandra in Low-Batch Inference}
Figure~\ref{fig_6_1} shows that Cassandra achieves up to a 2.41x speedup compared to the BFloat16 baseline. In contrast, 8-bit quantization yields only a 1.3$\times$ performance improvement over BFloat16. The observed performance degradation stems from the overhead associated with online activation quantization and scaling factor multiplication. In the prefill stage, where GEMM execution time is the main bottleneck, this overhead is largely masked by the gains from reduced GEMM computation. However, in low-batch LLM inference, the decode stage accounts for the majority of the end-to-end latency. Since GEMM is not the bottleneck in the decode stage, the overhead from activation quantization and scaling become non-negligible. Previous studies~\cite{Smoothquant, FP8-Quantization-Performance} have also reported that the decode stage performance of INT8 and FP8 quantization ranges from 1.25$\times$ to 1.42$\times$ compared to FP16 and BF16 baseline, which aligns with our measurements.

\subsubsection{Compatibility with Quantization}
While we utilize BFloat16 as our baseline, Cassandra is fully compatible with quantization. The most straightforward format to integrate into Cassandra is MXINT8. Currently, Cassandra-2 employs the MXINT format during the draft model generation process. Extending this approach to ensure the entire target model utilizes the MXINT8 format can be achieved seamlessly. According to prior research~\cite{INT-vs-FP}, MXINT8 exhibits overhead similar to other 8-bit quantization methods and generally offers superior accuracy compared to MXFP8.

Furthermore, with minor modifications, Cassandra can be extended to INT precision. Any-precision LLM~\cite{Any-Precision-LLM} serves as an excellent example of INT quantization that can be integrated with Cassandra. These methods propose multi-precision quantization techniques that allow for the deployment of optimized lower-bit models via simple truncation, which aligns perfectly with Cassandra’s core algorithm of generating draft models through pruning and truncation.

\subsection{Comparison with Other Speculative Decoding Methods}
\label{section_7_2}

\begin{figure}[t]
\centering
\includegraphics[width=\colimgwidth]{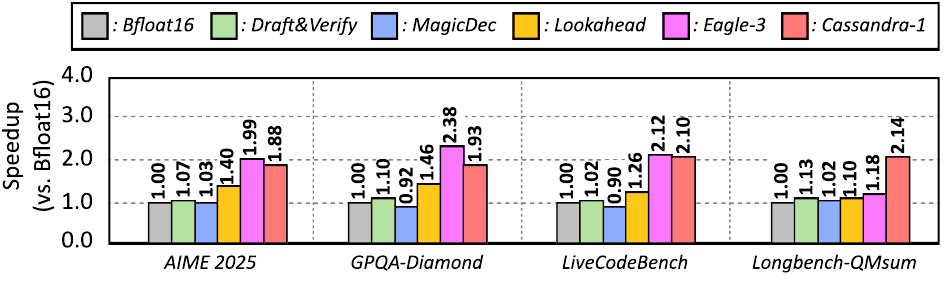}
\caption{Performance Comparison of Different Speculative Decodings.}
\label{fig_7_1}
\vspace{-1.5em}
\end{figure}

\subsubsection{Performance}

Figure~\ref{fig_7_1} illustrates the performance improvements of various speculative decoding schemes relative to the BFloat16 baseline. For our evaluation, we employed Draft\&Verify~\cite{Draft&Verify}, MagicDec~\cite{MagicDec}, and Lookahead Decoding~\cite{Lookahead-Decoding} as representative training-free speculative decoding methods, while EAGLE-3~\cite{eagle3} was utilized as the training-based speculative decoding baseline. All experiments were conducted using official open-source implementations. Also, we utilized DeepSeek-distill-Llama-8B~\cite{Deepseek-R1-Distillated-Llama3-8B} to leverage the official EAGLE-3 draft weights. Furthermore, since all speculative decoding frameworks utilize a BFloat16 target model, we selected Cassandra-1 as our baseline to ensure a fair comparison.

As shown in Figure~\ref{fig_7_1}, Cassandra consistently outperforms Draft\&Verify, MagicDec, and Lookahead Decoding across all four benchmarks. In the case of MagicDec, its reliance on KV cache pruning leads to significantly degraded performance in low-batch inference scenarios, occasionally even performing slower than the baseline.

Draft\&Verify also fails to generate a efficient draft model through its Bayesian-based layer skipping; although it skips 18 attention layers from the original 32-block model, it only skips 9 FFN layers. Consequently, the draft model must still load 70.7\% of the original model's parameters, a structure that is fundamentally limited in achieving high-speed execution for low-batch LLM inference.

Regarding Lookahead Decoding, which predicts subsequent tokens by referencing N-gram sets generated from previous data, it exhibited lower performance gains compared to Cassandra, particularly on LongBench and LiveCodeBench. Nevertheless, it achieved notable speedups of 1.40$\times$ and 1.46$\times$ on AIME2025 and GPQA-Diamond, respectively.

While EAGLE-3 demonstrated superior performance over Cassandra on AIME2025 and GPQA-Diamond, its effectiveness is heavily biased toward the characteristics of its training data. As illustrated, EAGLE-3 shows a significantly diminished margin of improvement in specific tasks such as long-sequence understanding. These results confirm that Cassandra not only outperforms most existing speculative decoding schemes across diverse scenarios but also provides robust performance gains with minimal sensitivity to sequence length or task type.

\subsubsection{Pre-Computation Cost}

To utilize Cassandra, two distinct pre-computation steps are required. The first step involves calibration for Wanda-based weight pruning. A small calibration set of approximately 128 samples is sufficient for this pruning method to deliver robust performance~\cite{Wanda}. Other techniques employed in Cassandra, such as exponent compression, KV cache pruning, and mantissa truncation, do not require any calibration. The second step is the optimization of ratio for pruning and truncation. The optimal combination of pruning ratio and truncation bits for configuring the draft model can be determined based on the acceptance rate relative to the compression ratio, which can be expressed by the following objective function.

\begin{equation}
\text{\small} 
\mathcal{J} = \frac{\alpha}{S_w(1 - w_p)(B - w_t) + S_{kv}(1 - kv_p)(B - kv_t)}
\end{equation}
\vspace{0.03in}

Here, $S_w$ denotes the weight size, $S_{kv}$ is the KV cache size, $w_t$ and $kv_t$ represent the truncation bits for weights and the KV cache respectively, $w_p$ and $kv_p$ are the pruning ratios for weights and the KV cache, $B$ is the number of bits used by the target model, and $\alpha$ is the acceptance rate.

As a practical approach to identifying a local optimum, we recommend prioritizing the optimization of hyperparameters associated with the dominant term by comparing the magnitudes of $S_w$ and $S_{kv}$. We conducted a grid search by incrementing the pruning ratio for weights and KV cache from 30\% to 60\% in 10\% intervals and the truncation range from 0 to 5 bits in 1 bit intervals. Following the experimental setup of Draft\&Verify, we used a development set of 8 samples. When using an 8B model, this process requires approximately 5 minutes of GPU time on an NVIDIA A100. In comparison with competing approaches, this overhead is acceptable; for instance, Eagle-3 requires 96 to 192 GPU hours on an A100, and Draft\&Verify requires approximately 1.5 hours of Bayesian optimization on A100 for an 8B model.


Additionally, unlike other hyperparameters that vary drastically across models, our default configuration, a 40\% pruning ratio and 4 bit truncation, demonstrates robust transferability to other models. Users can either adopt this configuration directly or use it as a starting point to narrow the search space, further reducing the cost of finding a local optimal solution.

\begin{figure}[t]
\centering
\includegraphics[width=\colimgwidth]{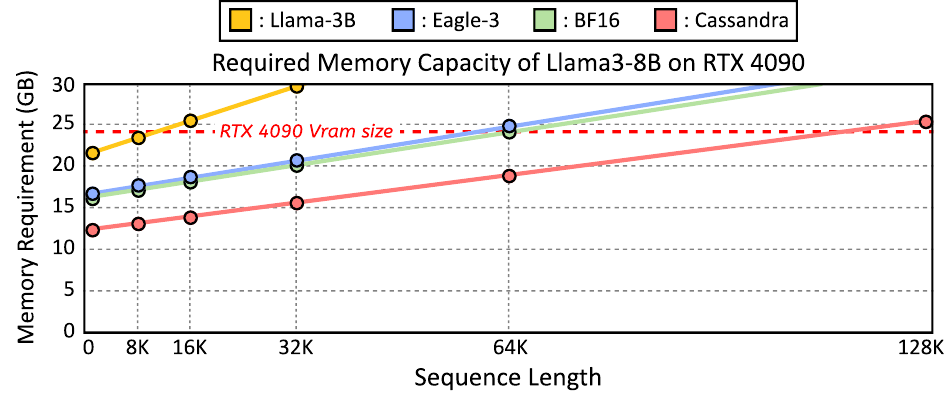}
\caption{Comparison of memory requirements between autoregressive decoding and various speculative decoding methods.}
\label{fig_6_3}
\vspace{-1.5em}
\end{figure}

\subsubsection{Memory Capacity Requirement}
Figure~\ref{fig_6_3} shows the ideal memory capacity requirement in various decoding schemes. As shown in the figure, Cassandra exhibits the most superior memory capacity efficiency compared to other speculative decoding methods and can generate 11.59$\times$ and 1.81$\times$ more tokens than Llama3-based speculative decoding and Eagle-3, respectively. Due to its exponent compression and memory-efficient draft model design, Cassandra actually requires less memory capacity than the original BFloat16 format. This is yet another reason why Cassandra is suitable for resource-constrained devices.

%% file: Contents/8_Related_Work.tex
\section{Related Work}
\label{section_8_1}

\textbf{Self-Speculative Decoding.} To address the limitations of training-based speculative decoding, a number of prior works have explored training-free approaches. Draft\&Verify~\cite{Draft&Verify} and Swift~\cite{swift_specdec} construct the draft model by skipping a subset of layers from the target model. In contrast, MagicDec~\cite{MagicDec} and QuantSpec~\cite{QuantSpec} construct the draft model by simplifying the KV cache of the target model via pruning and reduced-precision quantization, respectively. In some cases, these approaches additionally introduce separate weights for the draft model. Despite the diversity of self-speculative decoding techniques, their performance gains remain limited in low-batch scenarios. Moreover, when these methods introduce separate weights for the draft model, they incur non-negligible memory capacity overhead, which can be particularly problematic for consumer-grade devices.

\textbf{Systems for Speculative Decoding.} Recent research has actively explored hardware and system designs to accelerate speculative decoding. These works typically assume the existence of distinct target and draft models, and propose heterogeneous architectures tailored to their differing workload characteristics. For instance, some studies~\cite{SpecPIM, LP-spec, ADAPTIVE_DRAFT_LENGTH} employ processing-in-memory (PIM) devices to accelerate the draft model, while others, such as DFVG~\cite{Draft_FPGA_DFVG}, utilize FPGA-based acceleration. While such heterogeneous systems are effective for conventional training-based speculative decoding, they are not well-suited for self-speculative decoding, where the draft model is derived as a subset of the target model.

\textbf{Hardware Accelerators for Format Conversion.} Several prior studies~\cite{ecco, oaken, Deca} have investigated augmenting xPU systems with dedicated encoder and decoder units to support data compression and format transformation. For example, Ecco~\cite{ecco} proposes encoder and decoder designs for GPU-oriented cache compression, while Oaken~\cite{oaken} introduces similar mechanisms for KV cache compression. Unlike these prior efforts, which primarily focus on improving the efficiency of lossy compression, Cassandra proposes encoder and decoder architectures specifically designed to support speculative decoding.

%% file: Contents/9_Conclusion.tex
\section{Conclusion}
\label{section_9_1}
The availability of lossless LLM acceleration on consumer-grade devices has been notably limited to date. However, to overcome the inherent limitations of lossy compression that have emerged in the era of reasoning LLMs, we propose Cassandra, a training-free self-speculative decoding method tailored for low-batch LLM acceleration. Cassandra introduces a hardware-assisted approach that constructs a high-performance draft model based on fine-grained criteria without requiring additional training. Our experimental results demonstrate that this approach achieves superior throughput and significantly reduces the memory footprint compared to the baseline. This outcome suggests the unique advantage of Cassandra in fully harnessing the potential of reasoning LLMs within edge environments.